# Incoherent Quasielastic Neutron Scattering Study of Molecular dynamics of 4-n-cyano-4'-octylbiphenyl.

**Ronan Lefort*[a], Denis Morineau[a], Régis Guégan[b], Claude Ecolivet[a], Mohammed Guendouz[c], Jean-Marc Zanotti[d] and Bernhard Frick[e]**

*a Institut de Physique, UMR-CNRS 6251, Campus de Beaulieu, 35042 Rennes cedex, France. Fax: +(33)2 23 23 67 17; E-mail: ronan.lefort@univ-rennes1.fr*
*b Institut des sciences de la Terre, 1A, rue de la Férollerie, 45071 Orléans cedex 2, France*
*c Laboratoire d'Optronique, FOTON, CNRS-UMR 6082, Université de Rennes 1, F-22302 Lannion Cedex, France*
*d Laboratoire Léon Brillouin (CEA-CNRS), F-91191 Gif-sur-Yvette Cedex, France*
*e Institut Laüe-Langevin, 6 rue Jules Horowitz, F-38042 Grenoble, Cedex 9, France*

*†Electronic Supplementary Information (ESI) available:. See http://perso.univ-rennes1.fr/ronan.lefort*

We report incoherent quasielastic neutron scattering experiments on the thermotropic liquid crystal 4-n-cyano-4'-octylbiphenyl. The combination of time-of-flight and backscattering data allows analyzing the intermediate scattering function over about three decades of relaxation times. Translational diffusion and uniaxial molecular rotations are clearly identified as the major relaxation processes in respectively the nanosecond and picosecond time scales. The comparison with literature data obtained by other techniques is discussed.

At present, some most challenging issues are related to the physical properties of interfacial and nanoconfined fluids, with a special interest towards the structural relaxation processes (with its cooperative character[1]) and the glassy dynamics of confined liquids[2, 3].Whereas the case of simple liquids has been extensively investigated[3], many aspects of the dynamical behaviour of more complex nanoconfined fluids, like binary systems or anisotropic phases, still remain up to now unclear. Progresses in this field require a prior comprehensive characterization of the microscopic relaxation processes that occur in the bulk system. Very recently, the question of the universality of the structural relaxation of supercooled liquids and isotropic liquid crystals has been raised, and a unified description in terms of a hierarchical sequence of local and collective processes has been proposed[4, 5]. Despite their apparent higher level of complexity, mesogenic fluids can therefore be considered as exceptional reference systems in order to investigate the behaviour of individual molecular relaxation processes as a function of temperature, external field or confinement.

Cyanobiphenyls (nCB) are liquid crystals which physico-chemical properties have been extensively studied, partly because several members of this family can develop nematic phases at room temperature, and enter the composition of displays, or electro-optic devices[6]. Within this series, 4-n-cyano-4'-octylbiphenyl (8CB) holds a particular place, as it presents archetype isotropic to nematic and nematic to smectic A phase transitions. These remarkable features have been used in many experimental studies as reference observables, and followed during the application of external parameters such as spatial microconfinement[7-9], surface interaction[10, 11], or magnetic fields[12]. In particular, the dynamic behaviour of nanoconfined 8CB has recently raised a considerable interest[13-16], and a very special attention is paid to the understanding of quenched disorder effects on the collective and local molecular dynamics[17-23].

All these studies rely however only on a fractional knowledge of the actual molecular dynamics of unconfined 8CB. Whereas several studies like dielectric spectroscopy[14, 24, 25] or light scattering[26, 27] have focused on reorientational molecular modes, the translational self-diffusion tensor could be reported by resonance techniques[28]. However, these techniques are mode selective, and the time windows they can access differ by orders of magnitude. As a consequence, it remains difficult to bridge all pieces of information, and make sure that every essential feature of the liquid crystal relaxation is discovered. Surprisingly, very few incoherent quasielastic neutron scattering (IQNS) results were reported on cyanobiphenyls[29, 30], whereas numerous data already exist for many other liquid crystals[31-38].

The aim of this article is to provide a consistent picture of the microscopic relaxation modes, which contribute to the neutron intermediate relaxation function of bulk 8CB. An analytical model is built in order to disentangle the different active degrees of freedom and extract a limited numbers of relevant parameters. This comprehensive description should provide a valuable starting point towards the understanding of the origin of the very different microscopic dynamics of 8CB in complex environments reported recently.

The analysis by IQNS of the molecular dynamics of 8CB is performed from the picosecond to the nanosecond time scale in its isotropic and smectic A phases, the manuscript being organized as follows: after a brief description of the experimental procedure, we discuss the information obtained by high resolution backscattering IQNS. In a second part, it is completed at short times by time-of-flight data, in order to unravel the nature of the fast molecular dynamics by analyzing the full intermediate scattering function over more than three decades of correlation times.

## Experimental and technical details

### Samples and Neutron scattering experiments

Fully hydrogenated 8CB was purchased from Aldrich and used without further purification. It undergoes with increasing temperature the following sequence of phases: crystal (*K*), smectic A (*A*), nematic (*N*), and isotropic (*I*) with the transition temperatures: $T_{KA}$ = 294.4 K, $T_{NA}$ = 305.8 K, and $T_{NI}$ = 313.5 K.

Quasielastic neutron scattering experiments were carried out on the time of flight (TOF) spectrometer MIBEMOL at the Laboratoire Léon Brillouin (LLB) and the high resolution backscattering (BS) spectrometer IN16 at the Institut Laüe Langevin (Grenoble). The contribution from the incoherent cross section of the hydrogen atoms corresponds to 89% of the total scattering cross section of one 8CB molecule, so that the contribution to the measured scattering intensity from other atoms and from diffuse coherent scattering can be neglected within a good approximation. The coherent contribution associated to the Bragg reflection in the smectic A phase arises at Q=0.19 $Å^{-1}$. In order to avoid contamination of the incoherent data by this smectic peak, only the Q range above 0.5 $Å^{-1}$ was retained in the IQNS experiments. The time of flight MIBEMOL spectrometer allows one to probe a large energy transfer range with a relaxed resolution (FWHM) of 107 μeV at the elastic position for an incident wavelength of 6 Å. The elastic *q* range covered by MIBEMOL in this configuration extends from 0.45 to 1.95 $Å^{-1}$. The typical time window probed by these TOF experiments was about 1 to 20 ps. A standard configuration of the IN16 spectrometer was chosen with Si (111) monochromator and analyzers in backscattering geometry, which corresponds to an incident wavelength of 6.271 Å and results in a full width at half maximum (FWHM) energy resolution of 0.9 μeV. The dynamical range is ±15 μeV with a *q* range between 0.2 and 1.9 $Å^{-1}$, corresponding to a typical time window of about 1 to 10 ns. A cryoloop and a cryofurnace were respectively used on MIBEMOL and IN16 spectrometers in order to regulate the sample temperature in a range from respectively 100 to 340 K or 10 to 340 K. For the time-of-flight experiments, the sample holder was made of a folded 2 cm² aluminium sheet of slab geometry. For the backscattering experiments, the 8CB was deposited as a cylindrical film maintained in an ILL aluminium cell made of two interpenetrating hollow cylinders. In both cases, a large surface to volume ratio of the sample was completed, in order to maximize the matter volume in the neutron beam while minimizing multiple scattering. No macroscopic preferential orientation of the anisotropic phases of the liquid crystal was induced by the sample containers. The spectrometer resolution functions $R^{BS}(Q,\omega)$ for BS and $R^{TOF}(Q,\omega)$ for TOF were measured on the sample at 10K and 100 K, respectively.

### Brillouin scattering experiments

Brillouin scattering was performed with a triple pass tandem of Fabry–Perot (Sandercock model) with a $Kr^+$ ion laser (Coherent) at a wavelength of 647 nm. At too high counting rates a shutter obturates the photomultiplier inducing discontinuities in the spectra. Typical depolarized spectra of 8CB were accumulated over several hours, and reported elsewhere[20]. They do not show, as expected, any Brillouin doublet, but allow the best observation of the quasielastic region. They were fitted by an apparatus function convoluted to a lorentzian quasielastic line.

### IQNS data processing and fitting

Standard data corrections (empty cell substraction, self-absorption) were applied using conventional programs provided at ILL (SQW) and LLB (QENSH). TOF and BS data were partially averaged over groups of detectors following standard procedures in order to improve the final statistics. Consequently, same average values of the transfer vector of approximately Q = 0.5, 1 and 1.5 $Å^{-1}$ were retained for the two spectrometers. For TOF data, these three Q values were obtained by averaging spectra from $2\theta$ angles from respectively 23.5° to 27.5°, 52.6° to 57.6° and 84.5° to 95.5°. Within these groups, the energy dependence of the Q vector does not exceed 5% of the elastic value in the quasielastic region of interest, and interpolation of the data at constant Q is therefore unnecessary. For BS data, they were obtained by averaging spectra from respectively 25° to 38°, 57.5° to 70.5° and 90° to 103°. The resulting effective Q resolution was about $\Delta Q/Q_{elastic}$ = 10% for both TOF and BS.

The fitting of TOF and BS scattering functions $S(Q,\omega)$ in the frequency domain was completed using the programs provided respectively by the LLB (QENSH) and the ILL (online version of PROFIT). It was achieved using a standard empirical function made of the sum of one elastic peak (resolution limited) and several lorentzian lines. Time domain intermediate scattering functions $F(Q,t)$ at a given temperature were obtained by inverse Fourier transform (IFT) of the frequency domain data restrained to the quasielastic energy window (whole range $\pm E^{BS}_{window} = 15\mu eV$ for BS and $\pm E^{TOF}_{window}$ = 6 meV for TOF, constructed from the -6 to +1.4 meV data range combined with the symmetrization to positive energies of the -6 to -1.4 meV data range). The result of the IFT was deconvoluted from the apparatus resolution following equation 1 (*sp*=BS or TOF):

$$F^{sp}(Q,t) = \frac{IFT\left[S^{sp}(Q,\omega)\right]}{IFT\left[R^{sp}(Q,\omega)\right]} \quad (1)$$

All data points calculated for times larger than the spectrometers resolutions were discarded ($t > h/E^{sp}_{resol}$). Also short time cut-offs were used, discarding data points possibly artefacted by the truncation of the frequency domain data to the quasielastic region ($t < h/E^{sp}_{window}$)[39]. Following these filters, final time domain data extend from 0.6 to 20 ps for the TOF region (resp from 0.15 to 2 ns for the BS region).

## Results and discussion

### Frequency domain analysis of Backscattering data

Figure 1 shows typical BS spectra measured at T = 296 K in the smectic A phase of 8CB. All these spectra are significantly broader than the resolution of the spectrometer (no elastic peak is present). This feature was observed at all temperatures, either in the isotropic, nematic, or smectic A phase. This lack of elastic intensity is a direct proof that the molecular dynamics experiences a complete loss of correlation on the time and space scale of the experiment, and indicates molecular displacements larger than c.a. 15 Å on the nanosecond time scale. This non localized character of the dynamics, together with the increase of the spectral linewidth with Q, suggests that molecular self-diffusion is the dominant relaxation mechanism revealed by the BS experiment. If this was the only molecular relaxation mode present, the normalized incoherent scattering function would simply write:

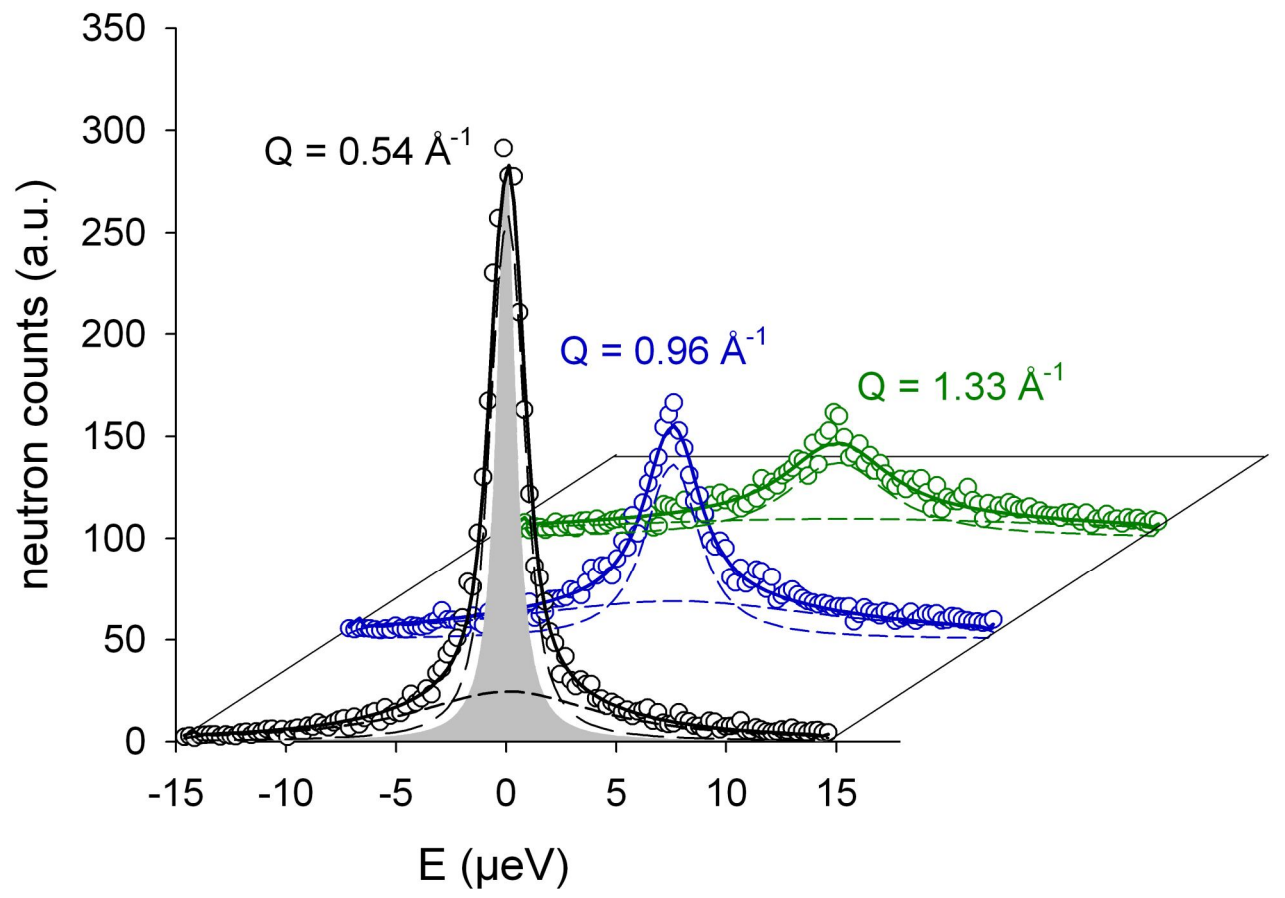

**Fig. 1** Incoherent quasielastic spectra of 8CB measured at T = 296 K on the high resolution backscattering spectrometer IN16 (ILL – Grenoble). The resolution function of the apparatus is displayed as a filled gray shape. The solid lines are best fits of the data (see text.) using a two components model (dashed lines).

$$S_{trans}^{BS}(Q,\omega)=\frac{1}{\pi}\frac{DQ^2}{\omega^2+(DQ^2)^2} \qquad (2)$$

This situation was indeed not observed for 8CB, for which a single lorentzian line was not enough to describe all BS spectra with a satisfactory agreement, except for temperatures above 310 K (isotropic phase). One possible origin of this non lorentzian shape would be the expected anisotropic character of the self diffusion[28, 40] occurring for $T < T_{NI}$. The anisotropic character of the diffusion is commonly quantified by the coefficient $\delta = (D_{//} - D_{\perp})/D_{\perp}$ , where the parallel symbol refer to the main principal axis of the diffusion tensor, often assumed to be aligned with the nematic director in case of liquid crystals. In that case, Dianoux *et al.* have shown[36] that the powder averaged quasielastic response of the anisotropic diffusion model all the more departs (is more peaked) from a pure lorentzian shape as the anisotropy coefficient $\delta$ is far from zero. All attempts to fit our data over the whole spectral range with this scattering law failed whatever the temperature, unless assuming an unphysical value of the anisotropy parameter close to $\delta = -1$, which would correspond to a purely bidimensional self-diffusion, either perpendicular to the director in the nematic phase, or confined within the smectic layers in the smectic A phase. To our knowledge, such 2D diffusion mechanism has never been observed for bulk nematogens, and would be in strong contradiction with the direct pulsed field gradient (PFG) NMR measurements of the molecular self-diffusion tensor of 8CB[40], which reported $+0.6 < \delta < +1$ on larger time (~ ms) and space (~ µm) scales than neutron scattering.

This gives strong indication for at least one additional and faster molecular relaxation process contributing to the BS window, leading to the non lorentzian shape of the incoherent scattering response of 8CB and making therefore very difficult to analyze the exact value of $\delta$ by IQNS. A most simple suitable model is constructed from the assumptions that : i) there is only one additional relaxation mode, ii) this mode is independent of the translational diffusion, iii) this mode has a localized character, then the incoherent scattering law can be classically approximated by the convolution product :

$$S_{total}^{BS}(Q,\omega) \propto S_{trans}^{BS}(Q,\omega) \otimes \left[A(Q)\delta(\omega) + (1-A(Q))\,Sq_{loc}^{BS}(Q,\omega)\right] \qquad (3)$$

In equation 3, the spatial restriction of the secondary motion gives rise to an elastic peak, which integrated intensity lead to the elastic incoherent structure factor (EISF) *A(Q)*. Its characteristic frequency is quantified by the linewidth $\Gamma_2(Q)$ of the quasielastic part $Sq^{BS}_{loc}(Q,\omega)$. An usual model free approach consists in assuming that this function is also a lorentzian line. Then equation 3 simply predicts the whole spectrum to be the sum of two lorentzians with integrated intensities $I_{trans}$ and $I_{loc}$ and with respective HWHM linewidths $\Gamma_1(Q)$ and $\Gamma_1(Q)+\Gamma_2(Q)$. The solid lines in figure 1 show the best fit of the data using equation 3 convoluted with the spectrometer resolution function. The agreement with the experiment is satisfactory owing to the statistical confidence interval of the neutron counts. Figure 2(a) displays the momentum transfer dependence of the linewidth $\Gamma_1(Q)$ of the thinnest line for three different temperatures at 315 K, 296 K and 280 K corresponding respectively to the isotropic, smectic and supercooled smectic phases. It shows a very clear linear behaviour with Q² over the whole Q range covered, in thorough agreement with equation 2. This is a direct proof that this line has to be assigned to a non restricted translational self-diffusion motion. From the slope of these curves, an effective diffusion coefficient *D* can be extracted, which values are plotted in figure 2 (b). They are in very good agreement with the isotropic average values $D_{iso}=(D_{//}+2D_{\perp})/3$ found by PFG NMR on a much larger lengthscale[40] (from 5 $10^{-11}$ m$^2$/s at 320 K to c.a. 6 $10^{-12}$ m$^2$/s at 280 K). This is a strong indication that the relaxation mechanism governing translational self-diffusion at the molecular level as measured by IQNS already holds the characteristic dynamic features that gives rise to the long range molecular transport on longer times scales. No discontinuity of *D* is detected within the accuracy of the IQNS experiment at the isotropic-nematic and neamtic-smectic phase transition temperatures, which is also in agreement with the temperature dependence of $D_{iso}$ as measured by PFG NMR. The activation barrier associated to translational self-diffusion can be determined with a good confidence from the temperature dependence of the integrated incoherent elastic intensity[19]. This can be achieved experimentally by switching off the Doppler machine of the BS spectrometer. In that case, only the elastic part $S^{BS}_{total}(Q,\omega=0)$ is measured, within the spectrometer energy resolution. Assuming the second process is significantly faster than the translational diffusion, then the condition $Sq^{BS}_{loc}(Q,\omega=0) << S^{BS}_{trans}(Q,\omega=0)$ is fulfilled, and equation 3 can be approximated by dropping the contribution of the second quasielastic term : $S^{BS}_{total}(Q,\omega=0) \approx A(Q)/(\pi DQ^2)$. If additionally this intensity is summed over all available detectors, the final measured value obeys equation 4,

$$I_{el}(T)=\int S^{BS}_{total}(Q,0)\,dQ \propto \frac{B}{D(T)} \qquad (4)$$

where *B* is temperature independent, which is expected if the geometry of the motion does not change with temperature. Figure 2 (c) shows the Arrhenius plot of $I_{el}(T)$, which remains fairly linear over 340 K down to crystallization that results in a sudden discontinuity of the data. From the slope of this graph, an activation energy of 38.3 kJ/mol is deduced, in perfect agreement with PFG NMR data[40] (also 38.3 kJ/mol).

Figure 3 shows the Q dependences of the linewidth $\Gamma_2(Q)$ of the second mode in the smectic phase of 8CB, as deduced from the comparison of the data with equation 3. In contrast with $\Gamma_1(Q)$, $\Gamma_2(Q)$ does not vanish for Q²→0, and displays only weak variations, often encountered for molecular rotational motions. For higher temperatures (nematic or isotropic phases), $\Gamma_2(Q)$ is not shown because its evaluation is questionable: $\Gamma_1(Q)$ becomes close to $\Gamma_2(Q)$ (5 to 6 µeV), which makes very difficult to disentangle the two modes..

An unambiguous assignment of this second quasielastic line to a given relaxation mechanism is hardly possible from BS experiments only. An extensive analysis of the relaxation processes up to higher frequency is required, as performed in the next section by the combination of BS and TOF.

Reference to previous IQNS studies conducted on a large variety of mesogenic phases[31, 34, 38, 41], including few neutron scattering experiments devoted to the family of cyanobiphenyls[29, 30], confirms the possible contribution of several relaxation mechanism to the IQNS spectra of 8CB. A general observation common to the nCB series is the occurrence of molecular dimerization at least in the nematic and smectic phases, leading to slower dynamics than other rod-like LCs, and to a pronounced translational-rotational decoupling[30, 31].

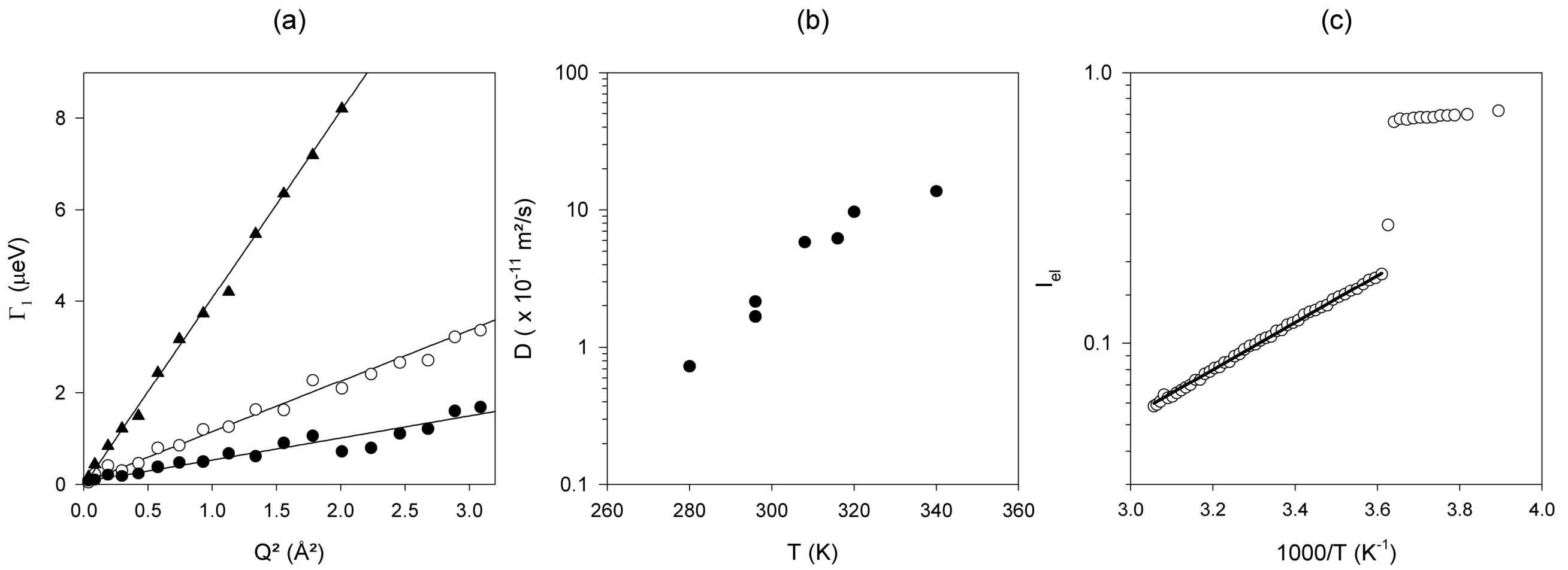

**Fig. 2** : Analysis of the slow component measured in the backscattering spectra of 8CB. (a) HWHM linewidth of the thinnest quasielastic line versus momentum transfer squared at 315K (triangles), 295K (open circles) and 280K (filled circles). The solid lines are best linear fits following $\Gamma_1 = D.Q^2$. (b) Effective self-diffusion coefficients determined from the linear fits of (a). (c) Arrhenius plot of the elastic intensity (resolution limited) integrated over all accessible Q values. The solid line is the best fit according the Arrhenius law $I_{el} \propto exp(-E_A/RT)$.

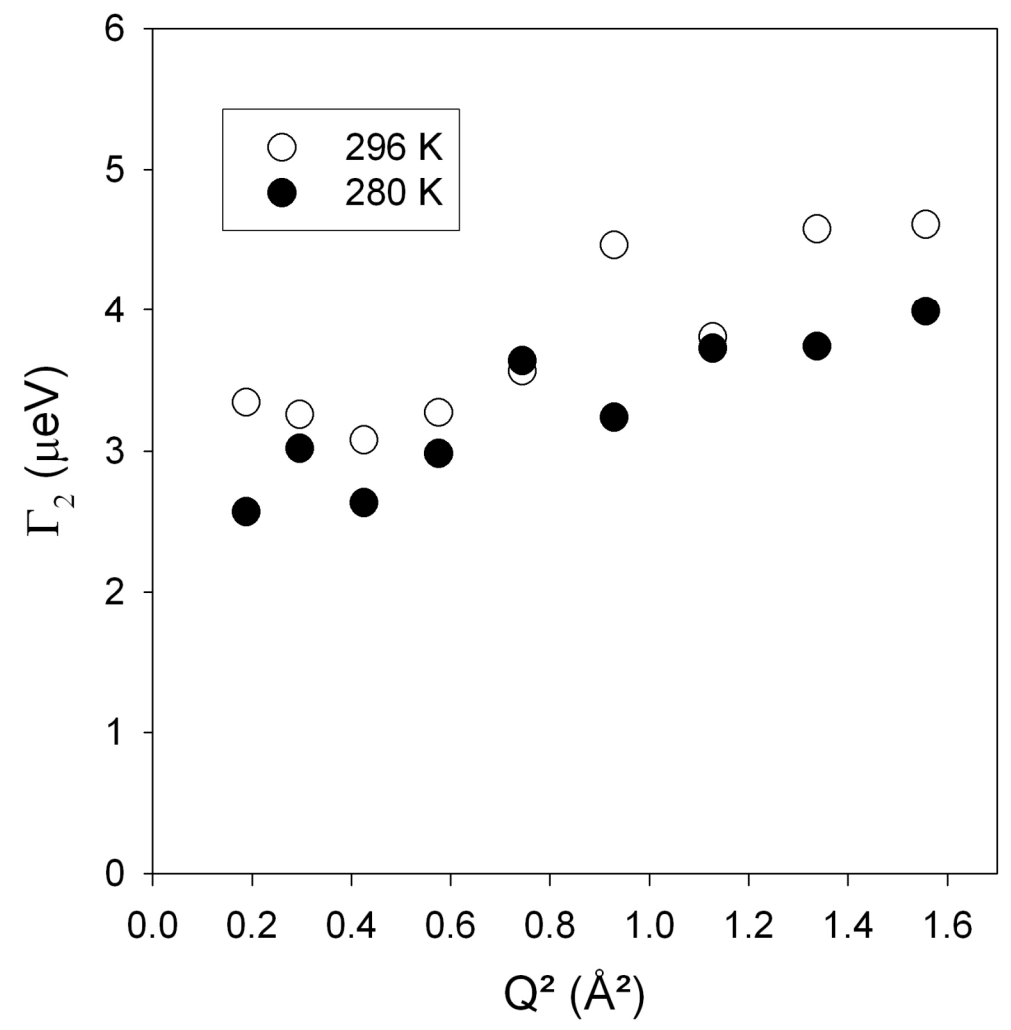

**Fig. 3** : HWHM linewidth $\Gamma_2(Q)$ of the fast mode measured in the backscattering spectra of 8CB at 296 K (open circles) and 280 K (filled circles).

Slow librational motions or flips around short molecular axis have been reported by dielectric relaxation experiments or analyzed by molecular dynamics simulations[42], with typical correlation times ranging from 1 to 10 ns. These are much slower than $\Gamma_2(Q)$ and are not likely to contribute to the BS spectra. On the contrary, chain-ends dynamics are expected to be much faster (around the picosecond time scale), and as it has been shown for other liquid crystals[33], they can be in a first approximation included in the Debye-Waller prefactor describing the small amplitude vibrational motions[33, 37]. Several optical techniques[26, 27] have reported that the molecular rotations along the 8CB long molecular axis occur with a characteristic time between 20 and 40 ps. This result was also confirmed in our group by Brillouin scattering experiments[20]. Such molecular reorientations are only 3 to 5 times faster than the BS window and could therefore contribute by the low frequency part of their spectral density.

### Time domain analysis of combined Time-of-flight and Backscattering data: the intermediate incoherent scattering function of 8CB

In order to combine experimental data acquired on different energy range, a time domain analysis has been preferred by calculating the intermediate scattering function $F^{sp}(Q,t)$ according to equation 1. This is a standard way to circumvent problems related to the convolution of the experimental spectra with very different energy resolution functions. The benefit of such a procedure has been proved by previous studies of the relaxation of polymers or glass-forming liquids[43-45] over several decades of relaxation times.

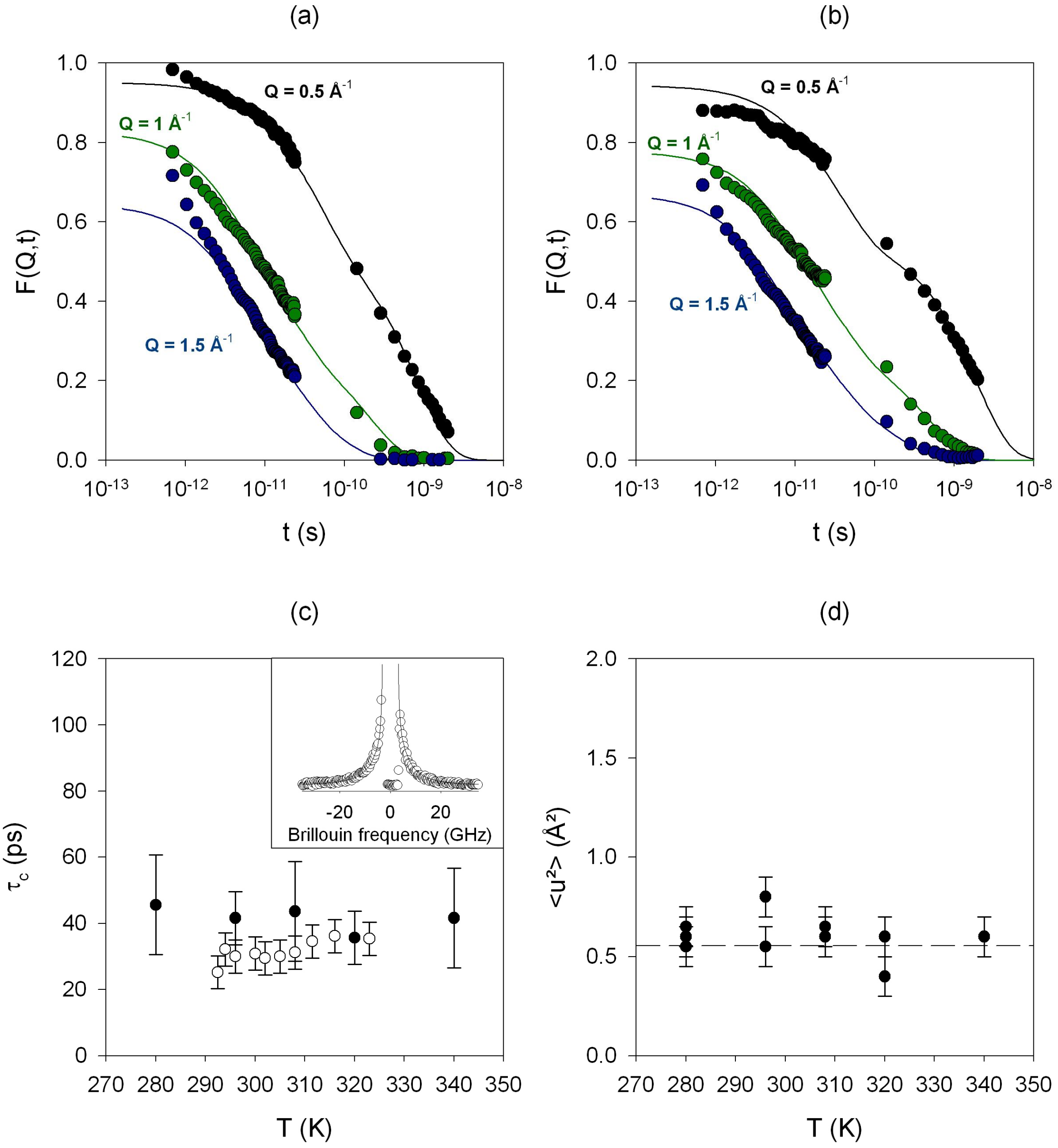

**Fig. 4 :** Time domain analysis of the intermediate scattering function of 8CB. (a) Intermediate scattering function calculated from combined TOF and BS data at T = 320 K in the isotropic phase of 8CB, at momentum transfer modulus of Q = 0.5, 1 and 1.5 $Å^{-1}$. The solid lines are best fits according to a model of motion including self-diffusion, uniaxial molecular rotation and fast vibrations (see text). (b) idem at T = 296 K. (c) Correlation time associated to the uniaxial rotation as deduced from the fit of the IQNS *F(Q,t)* (filled circles) and compared with the values obtained by Brillouin scattering experiments (open circles). The insert displays a typical Brillouin depolarized spectrum measured on bulk 8CB at 20°C. (d) Best fit values of the effective mean square displacement associated to the vibrational part of the model. For each temperature, the three values corresponding to the three values of Q displayed in (a) and (b) are reported.

Figure 4(a) and (b) displays the intermediate scattering functions of 8CB calculated from the combination of TOF and BS measurements at 320 K in the isotropic phase, and at 296 K in the smectic A phase. These relaxation functions decay to zero for sufficiently large values of Q. This complete loss of correlation is expected from the previous analysis of the BS data, as a consequence of self-diffusion. TOF and BS data do join together in a rather monotonous way, and do not reveal any clear multimodal decay. It is a strong indication that the slowest molecular modes that contribute to the TOF region cannot be faster than the self-diffusion process by much more than one decade. This is in full agreement with the conclusions based on the frequency domain analysis. Another important point is the strong Q dependence of the TOF data, that unambiguously indicates the presence of fast (~ ps) and large amplitude motions. These modes could be tentatively assigned to conformational degrees of freedom of the aliphatic chain extremities of the 8CB molecule. Indeed, fixed window scans have already revealed the existence of such mobility down to temperatures as low as 100 K (in the crystal phase), leading to an increased value of the mean squared displacement[19]. A stronger inflexion of the data at longer times suggest that the another relaxation mode can be suspected in the range from 10 to 100 ps. This is in agreement with the aforementioned uniaxial molecular rotation mechanism, as reported in this range by light scattering techniques. Based on these considerations, modelling of the intermediate scattering functions at different values of Q should therefore take into account at least three contributions: slow translational self-diffusion, uniaxial molecular rotations and large amplitude localized motions. The first mechanism has been very precisely analyzed in the frequency domain and can be therefore tightly constrained in a fitting procedure. The second one is a local process which is associated to an incoherent elastic contribution which structure factor highly depends on the molecular geometry (see Appendix). The most difficult task resides in the proper description of the large amplitude motions revealed by the TOF data. If they are identified to methyl groups rotations and librations of the chain ends, the exact fraction of hydrogen atoms implied in these motions and their correct geometry is not known. Hence, modelling these motions in a realistic way would require either to increase the number of free parameters, or to undertake more experiments on a liquid crystal with selectively deuterated chain ends. None of these procedures would however ensure to provide a more meaningful insight on the fast dynamics of 8CB. Instead, an oversimplified model was used, that identifies these chain end dynamics to vibrational-like motions which the typical amplitude is roughly measured by an effective mean square displacement $\langle u^2 \rangle$.

Following this approximation, a simple model of the intermediate incoherent scattering function can be built, and is presented in full detail in the appendix. In this model, the values of the self-diffusion coefficient $D$ are fixed to the one reported in figure 2(b), and only two independent parameters are then adjusted: the mean square displacement $\langle u^2 \rangle$ and the uniaxial rotation correlation time $\tau_c$. Both of these values were refined independently for each values of Q (and T), although the model predicts they should be Q independent. The results of these best fits are displayed as solid lines in figure 4(a) and (b), and the corresponding values of $\tau_c$ and $\langle u^2 \rangle$ are displayed in figure 4(c) and (d) respectively. In figure 4 (c), the average value of $\tau_c$ is shown (the values found at a given temperature for different Q vectors always lied in the domain specified by the error bars). The confidence interval for T = 280, 308 and 340 K is somewhat larger, because only BS data were available for these temperatures. The correlation times associated to $F^{rot}(Q,t)$ as deduced from our IQNS data are shown in figure 4(c) to be in very good agreement with those measured by Brillouin scattering[20]. This is of course an additional clue that the assignment of this mode to a molecular uniaxial rotation is correct. On the other hand, the model clearly fails to describe the data at very short times. This was indeed expected, due to the very crude approximation used to take chain end motions into account. This weakness can also be seen on figure 4(d), where the effective $\langle u^2 \rangle$ values are found very large, and their scattering around the average for certain temperatures are not fully consistent with the Q independence predicted by the model. Here again, this is a clear sign of the existence of large and fast amplitude motions contributing to the TOF window.

However, the overall agreement between the model and the experiment is still very satisfactory, as all essential features of the relaxation are quite well rendered over more than three decades of correlation times. It is noteworthy that this agreement could be reached even with only two free parameters and despite the strong constraints that were imposed on the motions geometry. This finally leads to a very simple physical picture of the molecular dynamics of 8CB in terms of elementary relaxation mechanisms, that is fully and quantitatively consistent with the results reported by light scattering or NMR experiments. Translation and rotation diffusion about the long axis are the dominant relaxation processes. Other relaxation mechanisms, especially "tumbling" or slow librations around short molecular axis, as reported by dielectric spectroscopy[14, 24] are probably too slow to contribute significantly to the IQNS intensity[29, 30]

## Conclusions

We reported in this paper incoherent neutron time-of-flight and backscattering data probing the molecular relaxation of the liquid crystal 8CB from the picosecond to the nanosecond time scale in its isotropic and smectic A phases. On the contrary to other techniques, IQNS on non oriented samples is not submitted to symmetry or polarization selection rules, and is able to explore in the same experiment all existing relaxation mechanisms. In spite of the resulting complexity of the measured dynamic cross-sections, we were able to analyze the molecular dynamics of 8CB within the frame of a very simple model including three elementary mechanisms: translational self-diffusion (nanosecond time scale), rotation around the long molecular axis (~10 ps timescale), and fast large amplitude motions (~ picosecond timescale) tentatively attributed to chain end librations and rotations. Here again, we find that this description of the relaxation function of the liquid crystal in terms of identified molecular motions is only possible because the different contributions, especially the rotational and translational parts, are well separated in time. One advantage of such description is to end up with a very simple analytical model able with a minimum set of parameters to reproduce the experimental observations, and that can be straightforwardly transferred to situations where the liquid crystal is submitted to external variables (confinement, solid-liquid interface, external field…), providing a new insight into the molecular dynamics of fluids in complex geometry.

## Appendix

Assuming that the molecular relaxation processes are independent vibrations, uniaxial rotation of the 8CB molecule around its long axis and molecular self-diffusion, the total incoherent intermediate scattering function writes:

$$F\left(Q,t\right) \approx \exp(-Q^2 \left\langle u^2 \right\rangle) F^{rot}\left(Q,t\right) \exp(-DQ^2 t) \quad (5)$$

In equation 5, $D$ is the self-diffusion coefficient and $F^{rot}(Q,t)$ is the intermediate scattering function for a uniaxial rotation.

Once properly averaged over the 8CB molecular geometry, $F^{rot}(Q,t)$ can be characterized by a single independent correlation time $\tau_c$. The intermediate scattering function of a single atom $i$ involved in random jumps between $N$ equivalent sites on a circle of radius $a_i$ is given by[41, 46] :

$$f_i^{rot}(Q,t) = A_i(Q) + \sum_{n=1}^{N-1} A_{i,n}(Q)\, e^{-t/\tau_n} \qquad (6)$$

where the EISF $A_i(Q)$ and the quasielastic structure factors $A_{i,n}(Q)$ are functions of the radius $a_i$ of the atom's trajectory:

$$A_{i,n}(Q) = \frac{1}{N}\sum_{k=1}^{N} J_0\left[2Qa_i \sin\left(\frac{k\pi}{N}\right)\right]\cos\left(\frac{2kn\pi}{N}\right) \qquad (7)$$
$$A_i(Q) = A_{i,0}(Q)$$

where $J_0(x)$ is the cylindrical Bessel function of the first kind. The total intermediate scattering function $F^{rot}(Q,t)$ can therefore be calculated by averaging these quantities over the radii $a$ associated to each of the $N_a$ hydrogen atoms of the 8CB molecule:

$$F^{rot}(Q,t) = \langle A_i(Q)\rangle + \sum_{n=1}^{N-1} \langle A_i(Q)\rangle_n\, e^{-t/\tau_n} \qquad (8)$$

This procedure allows all geometrical parameters to be fixed in the calculation. The $a_i$ radii were calculated using the molecular geometry of 8CB as reported in its crystal structure[47], as the distance between the position of the $i^{th}$ hydrogen atom and the long molecular axis. This axis is defined by the direction passing through the $C_2$ and $C_{11}$ carbon atoms of the phenyl rings, as illustrated in figure 5. It is likely that the radii for atoms close to the chain end are overestimated, as they probably experience additional large amplitude motions that could on the average bring them closer to the main rotational axis than in the crystal structure.

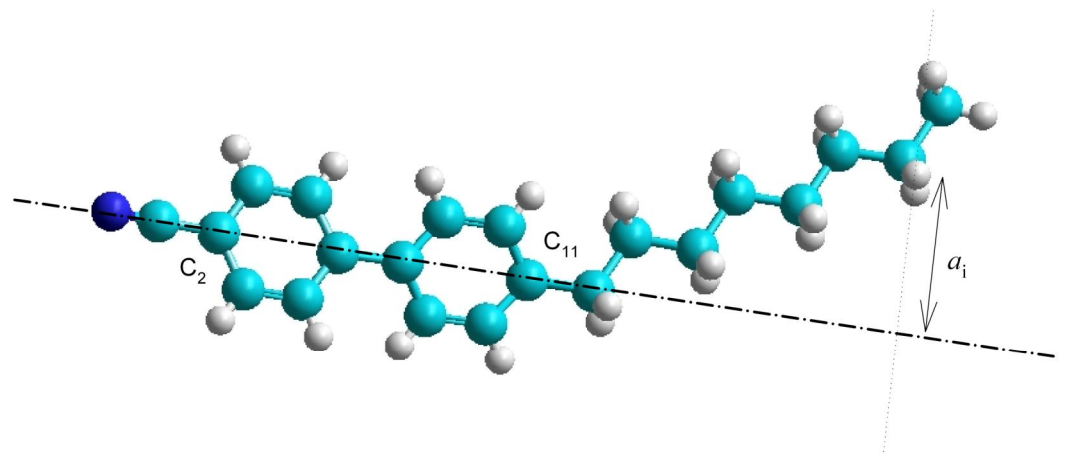

**Fig. 5 :** Definition of the long molecular axis of 8CB. The molecular geometry is taken from the crystal structure[47].

The correlation times $\tau_n$ appearing in the exponential series in equation 8 are related by:

$$\tau_n = \tau_c \frac{\sin^2\left(\dfrac{\pi}{N}\right)}{\sin^2\left(\dfrac{\pi n}{N}\right)} \qquad (9)$$

Hence, the number of free parameters implied in the model of uniaxial rotation of 8CB can be reduced to a single one ($\tau_c$) which gives a measure of the rotational diffusion constant ($D_{rot}$~$1/\tau_c$).

## Notes and references